\documentclass[twocolumn,showpacs,preprintnumbers,amsmath,amssymb,nofootinbib]{revtex4}
\usepackage{dcolumn}
\usepackage{bm}
\usepackage{tabularx}
\usepackage{array}
\usepackage{graphics}
\usepackage{graphicx}
\usepackage{psfrag}
\usepackage{epsfig}
\usepackage{amsmath}
\usepackage{amssymb}
\usepackage{verbatim}
\usepackage{color}
\usepackage{psfrag}
\usepackage{textcomp}
\usepackage[usenames,dvipsnames]{xcolor}

\usepackage{verbatim}

\newcommand{\msbar}{{\overline{\rm MS}}}

\newcommand{\nf}{N_{\rm f}}
\newcommand{\bea}{\begin{eqnarray}}
\newcommand{\eea}{\end{eqnarray}}
\newcommand{\beq}{\begin{equation}}
\newcommand{\eeq}{\end{equation}}
\newcommand{\gev}{{\rm GeV}}
\newcommand{\mev}{{\rm MeV}}

\newcommand{\pdir}{p\kern -5.2pt\raise 0.2ex\hbox {/}}
\newcommand{\vdir}{v\kern -5.75pt\raise 0.15ex\hbox {/}}
\newcommand{\kdir}{k\kern -5.75pt\raise 0.15ex\hbox {/}}
\newcommand{\epsdir}{\epsilon\kern -5.0pt\raise 0.15ex\hbox {/}}
\newcommand{\bvdir}{\bar{v}\kern -5.75pt\raise 0.15ex\hbox {/}}
\newcommand{\Ddir}{D\kern -7.75pt\raise 0.20ex\hbox {/}}
\newcommand{\Adir}{A\kern -7.75pt\raise 0.20ex\hbox {/}}
\newcommand{\ldir}{l\kern -5.0pt\raise 0.2ex\hbox{/}}
\newcommand{\varepsdir}{\varepsilon\kern -5.5pt\raise 0.15ex\hbox{/}}

\newcommand{\nn}{\nonumber}
\makeatother

\begin{document}
\preprint{\tt LPT Orsay 13-06}
\vspace*{22mm}
\title{On the significance of $B$-decays to the radially excited $D$ }

\author{Damir~Be\v{c}irevi\'c}
\affiliation{%
\vspace*{5mm}
Laboratoire de Physique Th\'eorique (B\^at. 210),\\ 
 Universit\'e Paris Sud,  Centre d'Orsay,
 91405 Orsay Cedex, France}

\author{Beno\^it~Blossier}
\affiliation{%
\vspace*{5mm}
Laboratoire de Physique Th\'eorique (B\^at. 210),\\ 
 Universit\'e Paris Sud,  Centre d'Orsay, 
 91405 Orsay-Cedex, France}

\author{Antoine~G\'erardin}
\affiliation{%
\vspace*{5mm}
Laboratoire de Physique Th\'eorique (B\^at. 210),\\ 
 Universit\'e Paris Sud,  Centre d'Orsay,
 91405 Orsay-Cedex, France}

\author{Alain~Le Yaouanc}
\affiliation{%
\vspace*{5mm}
Laboratoire de Physique Th\'eorique (B\^at. 210),\\ 
 Universit\'e Paris Sud,  Centre d'Orsay, 
 91405 Orsay-Cedex, France}

\author{Francesco~Sanfilippo} 
\affiliation{%
\vspace*{5mm}
Laboratoire de Physique Th\'eorique (B\^at. 210),\\ 
 Universit\'e Paris Sud,  Centre d'Orsay, 
 91405 Orsay-Cedex, France}

\date{\today}

\begin{abstract}
We discuss the possibilities of testing the recent claims of a relatively large $B\to D^\prime$ transition form factor through the non-leptonic $\overline B^0\to D^{\prime +}\pi^-$ and $B^-\to D^{\prime 0}\pi^-$ decay modes. To estimate the width of the latter we need the decay constant of the radially excited $D$-meson that we computed in lattice QCD with $\nf=2$ dynamical flavors and after taking the continuum limit we found $f_{D^\prime}=117(25)$~MeV. We also provide an update for the values of the $D$-meson decay constants obtained by using maximally twisted mass QCD. 
\end{abstract}

\maketitle

\section{\label{Introduction}Introduction}
Over the past several years there was a growing interest in radially excited $D$-mesons.  BaBar Collaboration has recently isolated a number of orbitally excited $D$-mesons, including a few states claimed to be the radial excitations~\cite{babar}. For our discussion particularly interesting is the $2S$ state, or $D^\prime$ meson, to which we focus in the following. From the angular distribution of the decay of $D^\prime \to D^{\ast}\pi$ the authors of ref.~\cite{babar} showed that the newly observed state is consistent with the $J^P=0^-$ assignment, and since its measured mass, $2539(8)$~MeV, turned out to be close to the quark model prediction  $m_{D^\prime}=2580$~MeV~\cite{gi}, they suggested the new state could be the $2S$-state or $D^\prime$. However, its measured width, $\Gamma(D^\prime)=130(18)$~MeV, appears to be much larger than expected in the quark models of ref.~\cite{swanson} which is why  the question of identification of the new state as a radial excitation remained open.~\footnote{For a recent review on the orbitally and radially charmed mesons, please see ref.~\cite{fulvia}.}   Note however that in quark models the transition matrix elements to radially excited states are hard to control because the position of the node in the wave function of the radial excitation highly depends on the type of equation used to derive wave functions. 

Besides the mass and width of the radial excitation it is particularly interesting to examine its weak interaction properties. Recently, in ref.~\cite{bernlochner}, it was suggested that a potentially large ${\cal B}(B\to D^\prime \ell \nu)$ could help solving the ``1/2 vs. 3/2 puzzle", because the subsequent $\Gamma(D^\prime\to D_{1/2}\pi)$ is much larger than $\Gamma(D^\prime \to D_{3/2}\pi)$ due to the fact that the emerging pion is in its $s$- and $d$-wave respectively. A large ${\cal B}(B\to D^\prime \ell \nu)$ would therefore result in an excess of the detected $B\to D_{1/2}(\pi)\ell \nu$ with respect to $B\to D_{3/2}(\pi)\ell \nu$. In a simplified notation, by $D_{1/2}$ we labelled a state belonging to the $j^P=(1/2)^+$ doublet of mesons $[D_0^\ast, D_1^\prime]$, while $D_{3/2}$ stands for a meson from the $j^P=(3/2)^+$ doublet  $[D_1, D_2^\ast ]$. The ``1/2 vs. 3/2 puzzle" refers to the fact that the $B$-factories observed  ${\cal B}(B \to D_{1/2}\ell\nu) \approx {\cal B}(B \to D_{3/2}\ell\nu)$  which turned out to be in conflict with predictions ${\cal B}(B \to D_{1/2}\ell\nu) \ll{\cal B}(B \to D_{3/2}\ell\nu)$~\cite{morenas,leibovich,puzzle1,puzzle2}. An argument in favor of significantly large $B\to D^\prime$ transition form factor has also been advanced in ref.~\cite{uraltsev} as it would provide a part of the suppression to the $B\to D^{(\ast)}$ form factors that could then explain the tension between $\vert V_{cb}\vert$, Cabibbo--Kobayashi-Maskawa (CKM) coupling, extracted from the inclusive and that extracted from the exclusive semileptonic decays.

In this paper we discuss the possibility to experimentally test the recent claims of  si\-gni\-ficantly large $\Gamma(B\to D^\prime \ell \nu)$ by measuring the corresponding non-leptonic decays. That possibility was first mentioned in ref.~\cite{bernlochner} which we further develop here. 
By using the factorization approximation, which appears to be very well verified by the data in the case of the so-called Class-I decays~\cite{neubert}, 
one can compare ${\cal B}(B\to D\pi)$ with ${\cal B}(B\to D^\prime \pi)$ and deduce 
\bea\label{eq:C1R}
{{\cal B}(\overline B^0\to D^{\prime +}\pi^-)\over {\cal B}(\overline B^0\to D^{+}\pi^-)} =(1.6\pm 0.1)\times |f_+^R(0)|^2\,,
\eea
so that for a large value of the $B\to D^\prime$ form factor, $f_0^{R}(m_\pi^2)\approx f_0^{R}(0)=f_+^{R}(0)$, one would have a large number of events that should 
be detectable in the data-sample accumulated at BaBar and Belle, and especially in that of LHCb. 
Furthermore, by means of numerical simulations of QCD on the lattice (with $\nf=2$ dynamical light quarks)  we computed the decay constant $f_{D^\prime}$ and obtained 
\bea
{f_{D^\prime}\over f_D}=0.57\pm 0.16\,,
\eea
so that the ratio between the Class-III and Class-I decays becomes 
\bea\label{eq:C3R}
{{\cal B}( B^-\to D^{\prime 0}\pi^-)\over {\cal B}(\overline B^0\to  D^{\prime +}\pi^-)} &={\displaystyle{\tau_{B^-}\over \tau_{\bar B^0}}} \left[ 1 + \displaystyle{0.13(4)\over f_+^R(0)} \right]^2\,.
\eea
The above estimates are made by assuming that the radial excitation is indeed the state observed by BaBar, $m_{D^\prime}=2539(8)$~MeV. As we shall see, the value of $m_{D^\prime}$ computed on the lattice is larger. That however  results in only slightly modified numbers on the right hand side (r.h.s.) of eqs.~(\ref{eq:C1R},\ref{eq:C3R}), and our statement that the $B\to D^\prime \pi$ decays should be detectable in the $B$-experiments for non-negligible $f_+^{R}(0)$ remains valid.

As a by-product of this study we improved the estimate of the $D$-meson decay constants made by the European Twisted Mass Collaboration (ETMC)~\cite{etmc-D,etmc-B} to
\bea
f_{D_s}=252\pm 3\ \mev , \quad {f_{D_s}\over f_D}=1.23(1)(1)\,,
\eea 
which translates to $f_D=205(5)(2)$~MeV. With respect to the results of ref.[13] in which the correlation functions with local operators have been considered only, here we combine those correlators with other ones in which several levels of smearing of interpolating operators have been included. Furthermore, we show that the chiral extrapolation of the double ratio $(f_{D_s}/f_D)/(f_K/f_\pi)$ is more stable than the one for $f_{D_s}/f_D$, which then helps us reducing the dominant source of systematic uncertainty, namely that associated with the chiral extrapolation.

The remainder of this paper is organized as follows: in Sec.~\ref{sec:Definitions} we introduce the form factors, comment on their existing estimates and describe how they could be extracted from the non-leptonic decays; in Sec.~\ref{sec:Lattices} we discuss the lattice determination of several quantities including $f_{D^\prime}$ and $f_{D_s^\prime}$; in Sec.~\ref{sec:Pheno}  we complete the phenomenological discussion of Sec.~\ref{sec:Definitions} and finally summarize in Sec.~\ref{sec:Summary}.

\section{How large is the $B\to D^\prime$ form factor?\label{sec:Definitions}}

The weak interaction matrix element driving the $B\to D^\prime \ell\nu$ decay is conveniently expressed in terms of hadronic form factors $f_{0,+}^R(q^2)$ as 
\begin{align}\label{eq:1a}
\langle D^\prime(k) \vert \bar c \gamma_\mu b& \vert\bar B(p)\rangle =  {m_B^2 - m_{D^\prime}^2 \over q^2 } q_\mu  f_0^R(q^2) \nn\\
&+
\left( p_\mu+k_\mu - {m_B^2 - m_{D^\prime}^2 \over q^2 } q_\mu \right) f^R_+(q^2)\ ,
\end{align}
where $q=p-k$, and the superscript ``$R$" is used to distinguish $B\to D^\prime$ transition from the more familiar $B\to D$ weak decay form factors. At $q^2=0$ the two form factors are equal, $f_+^R(0)=f_0^R(0)$. Contribution of the form factor $f_0^R(q^2)$ to the differential decay rate of $B\to D^\prime \ell \nu$ in the Standard Model comes with a factor $\propto m_\ell^2$, due to helicity suppression, and for the case of $\ell = e,\mu$ it can be safely neglected.  

Form factors are nonperturbative quantities and for a decay to the radial excitation they are very hard to extract from numerical simulations of QCD on the lattice. Apart from an early attempt in ref.~\cite{hein}, there are no results concerning the form factors $f_{0,+}^R(q^2)$ reported so far. Instead, there are estimates made by using the constituent quark model~\cite{galkin,wang}.  An attempt to extract  $f_+^R(0)$ in the framework of QCD sum rules has been reported in ref.~\cite{bernlochner}. 

Similarly to $B\to D\ell \nu$ decay, also in this case the considerations in the heavy quark limit  are very useful. For $m_{c,b}\to \infty$ the form factor  $ f^R_+(q^2)$ is related to its Isgur-Wise function $\xi_R(w)$ as
\bea
{\sqrt{4 m_B m_{D^\prime}}\over m_B+m_{D^\prime}}\  f^R_+(q^2) \to  \xi_R(w)\,,
\eea
where $q^2$ and the relative velocity $w$ are related via $q^2=m_B^2+m_{D^\prime}^2-2m_Bm_{D^\prime}w$. Since the ground state and the radial excitation are orthogonal, the Isgur-Wise function at the zero recoil limit is $\xi_R(1)=0$. This is in contrast with the usual Isgur-Wise function for the elastic transitions~\cite{iwf} in which case $\xi(1)=1$, and $\xi(w)$ is then conveniently parametrized by the first few terms in expansion around the zero recoil limit,
\bea\label{eq:xi0}
\xi(w)=1 - \rho^2 (w-1) +\frac{1}{2}\sigma^2 (w-1)^2  \,.
\eea
Stringent bounds on the shape of the Isgur-Wise function $\xi(w)$ were derived in ref.~\cite{bounds}, of which the most interesting one for our purpose is the relation between the slope $\rho^2$ and curvature $\sigma^2$ of the ordinary Isgur-Wise function on one side, and the slope of the Isgur-Wise function relevant to the transition matrix element between the ground state and the radially excited one $\xi_R(w)$, namely~\footnote{Radiative corrections to the bounds of ref.~\cite{bounds} have been computed in heavy quark effective theory in ref.~\cite{dorsten} and are found to be very small.} 
\bea\label{eq:bound}
|\rho_R^2| \equiv \vert \xi_R^\prime(1)\vert \leq \sqrt{\frac{5}{3}\sigma^2 -\frac{4}{3}\rho^2 -(\rho^2)^2 }\,.
\eea
Experimenters fit their $B\to D\mu\nu$ data to the form~\cite{cln},
\bea
\xi(w) = {G(w)\over G(1)}= 1-8 \rho_D^2 z +(51\rho_D^2-10)z^2+\dots
\eea
where $z=(\sqrt{w+1}-\sqrt{2})/(\sqrt{w+1}+\sqrt{2})$, so that $\rho^2$ and $\sigma^2$ in eq.~(\ref{eq:xi0}) can be identified as
\bea
\rho^2=\rho_D^2,\quad \sigma^2={67\rho_D^2-10\over 32}\,,
\eea
which, together with the measured $\rho_D^2=1.19(4)(4)$~\cite{hfag}, inserted in eq.~(\ref{eq:bound}) gives
\bea
\vert \rho_R^2 \vert \leq 0.8\ .
\eea
Since $\xi_R(1)=0$ by definition, the only plausible way to enhance $f_+^R(q^2)$ is through large power corrections at the zero recoil point  ($w=1$), i.e. to $f_+^R(q^2_{\rm max})$ where  $q^2_{\rm max}=(m_B-m_{D^\prime})^2$. In this way the form factor would be significant in the low $q^2$ region too which due to the large phase space enhancement would provide a significant ${\cal B}(B\to D^\prime \ell\nu)$. 
The numerical estimates of this form factor have been made in the framework of two quark models, and both values are in fact large: $f_+^R(0)=0.37-0.41$~\cite{galkin}, and $f_+^R(0)\simeq 0.2$~\cite{wang}. A rough QCD sum rule estimate gives $f_+^R(0)=0.16(11)$~\cite{bernlochner}. 

\subsection{Non-leptonic decays can help}

From the above discussion we see that a large value of $f_+^R(q^2)$ at low $q^2$'s is difficult to reconcile with the bound~(\ref{eq:bound}) unless huge power corrections modify the heavy quark mass limit of the form factor. The size of $f_+^R(0)$ can be tested experimentally by considering the Class-I non-leptonic decays for which the factorization approximation is known to be very well verified in experiments and is formally proven to be valid in the heavy quark limit~\cite{fact,fact2}. More precisely, we will consider  $\overline B^0\to D^{\prime +}\pi^-$ decay for which the factorization  amplitude reads,
\begin{align}
{\cal A}^I_{\rm fact}=&-{G_F\over \sqrt{2}}V_{cb}V_{ud}^\ast a_1  \langle\pi^-\vert \bar d \gamma^\mu\gamma_5 u\vert 0\rangle  \langle\ D^{\prime +}\vert \bar c \gamma_\mu b\vert \overline B^0\rangle\nn\\
=&-i {G_F\over \sqrt{2}}V_{cb}V_{ud}^\ast a_1 f_\pi (m_B^2-m_{D^\prime}^2) f_0^R(m_\pi^2)\,,
\end{align}
where $a_1(m_b)$ is the Wilson coefficient that contains information about physics at short distances, and $f_0^R(m_\pi^2)\approx f_0^R(0)= f_+^R(0)$. When combined with the  more familiar $\overline B^0\to D^{ +}\pi^-$ decay mode, one gets
\begin{align}\label{eq:c1}
&{{\cal B}(\overline B^0\to D^{\prime +}\pi^-)\over {\cal B}(\overline B^0\to D^{+}\pi^-)} =
\left({m_B^2 -m_{D^\prime}^2\over m_B^2 -m_{D }^2}\right)^2  \times\hfill \nn\\
& \quad \quad \times  \left[{\lambda(m_B,m_{D^\prime},m_\pi)\over \lambda(m_B,m_{D},m_\pi)}\right]^{1/2} \left|{f_+^R(0)\over f_+(0)}\right|^2\,,
\end{align}
where $\lambda(x,y,z)=[x^2-(y+z)^2][x^2-(y-z)^2]$, and $f_+(0)$ is the $B\to D$ form factor, the value of which has been measured experimentally. In the lowest bin, $m_\mu^2\leq q^2\lesssim 1\ \gev^2$, the BaBar Collaboration obtained $\vert V_{cb}\vert f_+(q^2)\approx \vert V_{cb}\vert f_+(0)=0.02642(8)$~\cite{ffatzero}, which then, after using $\vert V_{cb}\vert = 0.0411(16)$~\cite{uta}, gives $f_+(0)=0.64(2)$. With this last number, and with $m_{D^\prime}=2.54$~GeV, eq.~(\ref{eq:c1}) gives
\bea\label{eq:CC}
{{\cal B}(\overline B^0\to D^{\prime +}\pi^-)\over {\cal B}(\overline B^0\to D^{+}\pi^-)} =(1.65\pm 0.13)\times |f_+^R(0)|^2\,.
\eea
$\overline B^0\to D^{+}\pi^-$ has been extensively studied at the $B$-factories, and the final result ${\cal B}(\overline B^0\to D^{+}\pi^-)=0.268(13)\%$~\cite{PDG} can be combined with eq.~(\ref{eq:CC}) to predict 
\begin{align}
{{\cal B}(\overline B^0\to D^{\prime +}\pi^-)} = |f_+^R(0)|^2 \times (4.7\pm 0.4)\times 10^{-3}\,,
\end{align}
which for the values of  $f_+^R(0)$ discussed in the literature leads to ${{\cal B}(\overline B^0\to D^{\prime +}\pi^-)} \sim 10^{-4}$, that should be possible to measure even in the sample of BaBar and Belle, and especially at LHCb. 

If the experimenters indeed detect a significant number of $\overline B^0\to D^{\prime +}\pi^-$ events then it would be interesting to check if the picture is consistent with what happens in the Class-III decay $B^-\to D^{\prime 0}\pi^-$. The factorized amplitude of this mode contains two pieces,
\begin{align}
{\cal A}^{III}_{\rm fact}=-&i {G_F\over \sqrt{2}}V_{cb}V_{ud}^\ast \biggl[ a_1 f_\pi (m_B^2-m_{D^\prime}^2) f_0^R(m_\pi^2) \biggr. \nn\\
& \biggl.  + a_2  f_{D^{\prime}}   (m_B^2-m_{\pi}^2) f_0^{B\to\pi}(m_{D^\prime}^2) \biggr]\,,
\end{align}
where $f_0^{B\to\pi}(m_{D^\prime}^2)$ is the $B\to \pi\ell \nu$ decay form factor. One could then consider,  
\begin{equation}\label{eq:three}
\begin{split} 
{{\cal B}( B^-\to D^{\prime 0}\pi^-)\over {\cal B}(\overline B^0\to  D^{\prime +}\pi^-)} &={\displaystyle{\tau_{B^-}\over \tau_{\bar B^0}}}\left[ 1 + {a_2\over a_1}\times {m_B^2-m_\pi^2\over m_B^2-m_{D^\prime}^2} \right. \cr
&\hspace*{-2mm}\left. \times  {f_0^{B\to \pi}(m_{D^\prime}^2)\over f_+^R(0)}\  {f_{D^\prime}\over f_D} \ {f_D\over f_\pi}\right]^2\,,
\end{split}
\end{equation}
in which two unknown quantities are $f_{D^\prime}/f_D$ and the ratio of the Wilson coefficients $a_2/a_1$. The first quantity can be determined on the lattice, while the second one can be extracted from the measured 
${\cal B}( B^-\to D^{0}\pi^-)=0.481(15)\%$ and the above-mentioned ${\cal B}(\overline B^0\to D^{+}\pi^-)=0.268(13)\%$~\cite{PDG}. 


 \section{Lattice determination of the decay constants $f_{D_{(s)}}$ and $f_{D^\prime_{(s)}}$~\label{sec:Lattices}}
In this section we present the results of our computation of masses and decay constants of $D^{(\prime)}$- and $D_s^{(\prime)}$-mesons by using the maximally twisted QCD (MtmQCD) action on the lattice~\cite{fr} with $\nf=2$ dynamical light flavors. At fixed lattice spacing, we will also compare the MtmQCD values with the results obtained by using the standard Wilson-Clover action with $\nf=2$ dynamical quarks and with those obtained in quenched QCD. Moreover, we will improve the values for $f_{D_{(s)}}$ and $f_{D_{(s)}}/f_D$ reported in refs.~\cite{etmc-D,etmc-B}.

Extraction of the mass and the decay constant of a given hadron state on the lattice is made from the study of the two-point correlation function,  
\begin{equation}\label{eq:two}
\begin{split} 
C_{ij}(t)&= \sum_{\vec x} \langle O_{\Gamma_i}(\vec  0,0)  O_{\Gamma_j}^\dag (\vec x, t) \rangle \\
& =-\langle \sum_{\vec x}{\rm Tr}\left[  \Gamma_i S_c(0,x) \Gamma_j S_q(x,0)  \right]\rangle \,, 
\end{split}
\end{equation}
where $O_{\Gamma_i}=\bar c\Gamma_i q$ is the bilinear quark operator, with $c$ and $q$ being the charm- and the light-quark field respectively, and $\Gamma_{i,j}$ is chosen to ensure the coupling to the state with desired quantum numbers. In our study $q$ is either the strange quark, or it coincides with the light sea quark.  In the above notation $S_q(x,0) \equiv \langle q(x)\bar q(0)\rangle$ is the quark propagator computed in the background gauge field configuration by inverting the Wilson-Dirac operator of MtmQCD on the lattice. The simplest and the most convenient choice of the operators is to use $O_{\Gamma_{i,j}}= P_5=  \bar c\gamma_5 q$, and extract the mass and decay constant of the lowest lying state from the exponential fall-off of~(\ref{eq:two}) which for large time separations,  
\begin{equation} \label{eq:2pts}
\begin{split} 
C_{55}(t)&= \langle {\displaystyle \sum_{\vec x} }  P_5(\vec x; t)  P_5^\dagger (0; 0) \rangle  \xrightarrow[]{\displaystyle{ t\gg 0}}  \\
 &\;  \left| {\cal Z}_{D_q} \right|^2 \frac{\cosh[  m_{D_q} (T/2-t)]}{ m_{D_q} } e^{- m_{D_q} T/2}\,,
\end{split}
\end{equation}
with $T$ being the size of the temporal extension of the lattice, and $ {\cal Z}_{D_q}  =\langle 0\vert \bar c\gamma_5 q\vert D_q\rangle$. In eq.~(\ref{eq:2pts}) we used the symmetry of the correlation functions with respect to $t\leftrightarrow T-t$ of our periodic lattice.
To extract the radial excitation properties one can subtract the r.h.s. of eq.~(\ref{eq:2pts}) from the correlator $C_{55}(t)$, 
\begin{align}\label{eq:subtr}
C_{55}^\prime &(t) = C_{55}(t) - \cr
& -\left| {\cal Z}_{D_q} \right|^2 \frac{\cosh[  m_{D_q} (T/2-t)]}{ m_{D_q} } e^{- m_{D_q} T/2},
\end{align}
and check whether or not there is a plateau of the effective mass, $m_{D_q^\prime}^{\rm eff}(t)$,  defined as
\bea\label{Meff}
{\cosh\left[ m_{D_q^\prime}^{\rm eff}(t) \left( {\displaystyle{T\over 2}} - t\right)\right] \over \cosh\left[ m_{D_q^\prime}^{\rm eff}(t) \left( {\displaystyle{T\over 2}} - t -1\right)\right] }  = {C_{55}^{\prime}(t)\over C_{55}^{\prime}(t+1)}\,,
\eea
and possibly fit to the form similar to eq.~(\ref{eq:2pts}) to extract the mass and the decay constant of $D^\prime_q$. This strategy can be extended and combined with correlation functions computed by using different source operators.

Another way to proceed is to work with several interpolating operators that can be easily built if instead of the local fields $q$ and $c$ one uses the smeared ones, $q_{n_g}$ and $c_{n_g}$, that we generically call $\psi_{n_g}$, defined via
\begin{equation}
\psi_{n_g}=\left(\frac{1+\kappa_g H}{1+6\kappa_g}\right)^{n_g}\psi\,,
\end{equation}
where the smearing operator $H$ reads~\cite{Gusken:1989ad}
\begin{equation}
H_{i,j}=\sum_{\mu=1}^3\left(U^{n_a}_{i;\mu}\delta_{i+\mu,j}+U^{n_a\dagger}_{i-\mu;\mu}\delta_{i-\mu,j}\right)\,,
\end{equation}
with $U^{n_a}_{i,\mu}$ being the $n_a$ times APE smeared link~\cite{Albanese:1987ds}, defined in terms of $(n_a-1)$ times smeared link $U^{(n_a-1)}_{i,\mu}$ and its surrounding staples  of links denoted by $V^{(n_a-1)}_{i,\mu}$, 
\begin{equation}
U^{n_a}_{i,\mu}={\rm Proj_{SU(3)}}\left[(1-\alpha)U^{(n_a-1)}_{i,\mu}+\frac{\alpha}{6} V^{(n_a-1)}_{i,\mu}\right ]\,.
\end{equation}
The above steps are known as the Gaussian smearing procedure. In this work, we choose the following values of the parameters:
\begin{equation}\label{eq:params}
\kappa_g=4,\,n_g\in (0,2,10,32),\,\alpha=0.5,\,n_a=20\,.
\end{equation} 
We checked that the correlation function computed with both quark fields smeared is equal to the one obtained with only one field smeared but with twice as many  smearing steps, $n_g$. However, we observe that the correlation functions computed with both fields smeared are less noisy and for that reason the results presented in this work are obtained by using both $q$ and $c$ fields smeared.  Therefore, with various values of $n_g$, and for $\Gamma=\gamma_5$ we get various operators $O_{i}$ that can be combined in the matrix of correlation functions~(\ref{eq:two}). Note that the choice $n_g=0$ corresponds to the local operator which is needed for the computation of physically relevant decay constants. The problem of extraction of the hadron masses from 
\begin{equation} 
\begin{split} 
&C_{ij}(t)=\langle O_i(t)O^\dag_j(0)\rangle  \cr
&= \sum_n \frac{{\cal Z}_{i,n} {\cal Z}^\ast_{j,n}}{m_{D_q^{(n)}}}
e^{-m_{D_q^{(n)}} \frac{T}{2}} \cosh\left[ m_{D_q^{(n)}}  \left({T\over 2}-t\right)\right],
\end{split}
\end{equation}
is then reduced to the generalized eigenvalue problem (GEVP)~\cite{GEVP}
\bea\label{eq:gevp}
&C_{ij}(t) v^{(n)}_j(t,t_0)=\lambda^{(n)}(t,t_0)
C_{ij}(t_0) v^{(n)}_j(t,t_0), \quad
\eea
where $C_{ij}(t_0)$ is chosen for computational (numerical) convenience, while $\lambda^{(n)}(t,t_0)$ and $v^{(n)}_j(t,t_0)$ are the eigenvalues and eigenfunctions of the matrix ${\bf C}^{-1}(t_0)  {\bf C}(t) $. The goal we achieve by solving GEVP is the identification of interpolating field  $\widetilde{O}^{(n)}\equiv v^{(n)}_i O_i$, which has a desired property that  
$\langle  D_q^{(m)} \vert \widetilde{O}^{(n)\dag}\vert 0\rangle =A_n \delta_{mn}$.  
Masses are then obtained from 
\bea\label{eq:spectral}
\lambda^{(n)}(t,t_0)={\cosh\left[ m_{D_q^{(n)}} (T/2-t)\right]\over \cosh\left[ m_{D_q^{(n)}} (T/2-t_0)\right]}\,,
\eea
where $n=1$ corresponds to the lowest lying pseudoscalar $D_q$ mesons, and $n=2$ to their first radial excitation $D_q^\prime$. 
To extract the decay constant one needs the matrix element of the local operator and a state $\vert D_q^{(n)}\rangle$ isolated by using $\widetilde{O}^{(n)}$. In other words, 
\bea\label{eq:me}
\langle  D_q^{(n)} \vert O_L^\dag \vert 0\rangle 
=
{\sqrt{A_n}~~\displaystyle \sum_i C_{Li}(t)v^{(n)}_i(t,t_0)\over \displaystyle  \sum_{ij}
v^{(n)}_i(t,t_0)C_{ij}(t)
v^{(n)}_j(t,t_0)}.~~
\eea
In the case of MtmQCD on the lattice, the local operator of interest is $ O_L= P_5=\bar c \gamma_5 q$ because $(\mu_q + \mu_c) P_5$ is renormalization group invariant, and therefore no renormalization constant is needed to compute the pseudoscalar decay constant.~\footnote{ $\mu_{q,c}$ is the quark mass parameter. } This is not so in the case of Wilson-Clover action where it is more convenient to use $O_L=A_0=Z_A(g_0^2) \bar c \gamma_0\gamma_5 q$, with $Z_A(g_0^2)$, the axial current renormalization constant. However, since we shall be interested in the ratio of the decay constants, $f_{D_q^\prime}/f_{D_q}$, one can use $ O_L= P_5$ in the case with the Wilson-Clover action as well.

\subsection{Lattices used in this work}

 \begin{figure*}[t!]
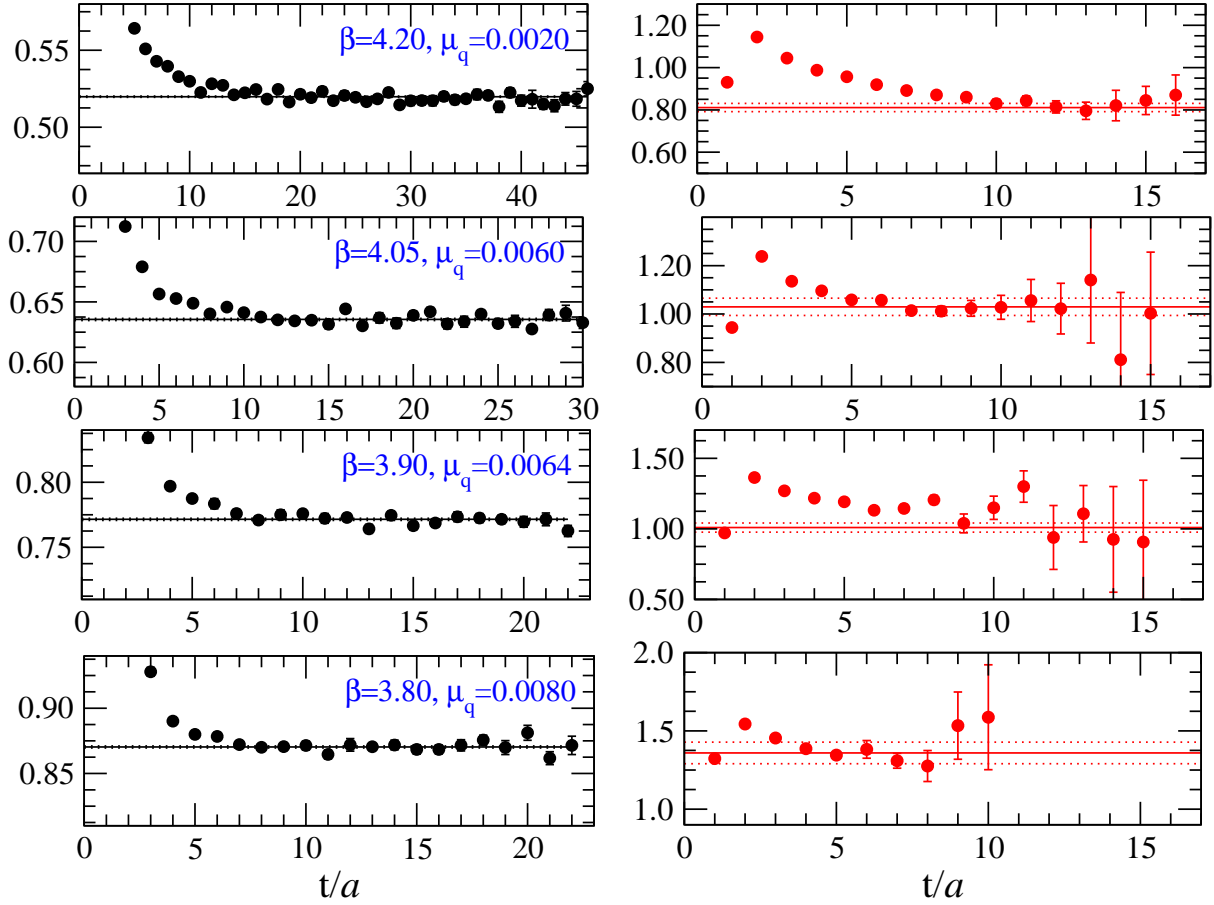

 \epsfig{file=420_0.eps, height=2.8cm} \quad\epsfig{file=420_1.eps, height=2.8cm}\\
 \epsfig{file=405_0.eps, height=2.8cm} \quad\epsfig{file=405_1.eps, height=2.8cm}\\
 \epsfig{file=390_0.eps, height=2.8cm} \quad\epsfig{file=390_1.eps, height=2.8cm}\\
 \epsfig{file=380_0.eps, height=3.3cm} \quad\epsfig{file=380_1.eps, height=3.47cm}\\
\caption{\label{fig:1}{\sl 
Effective mass plots leading to the masses of $D_s$ (left column) and $D_s^\prime$ states (right column), deduced from the matrix of correlation functions as discussed in the text. Plots are provided for all four lattice spacings considered in this work with MtmQCD action. } }
\end{figure*}
The lattice results of this paper are obtained by relying on the ensembles of gauge field configurations produced by the ETMC~\cite{Boucaud:2008xu} from the simulations of MtmQCD~\cite{fr} with $\nf=2$ dynamical mass degenerate light quarks. Main information concerning these ensembles is given in Tab.~\ref{tab:01}. We use the results of ref.~\cite{Blossier:2010cr} to fix the charm ($c$) and strange ($s$) quark masses at each lattice spacing and then compute the correlation functions needed for the extraction of strange and non-strange $D_q^{(n)}$-meson properties.  The quark propagators are computed by using stochastic sources, and in the computation of the correlation functions we used the so-called one-end trick~\cite{Boucaud:2008xu}.

The effective mass plots are obtained from corresponding eigenvalues $\lambda^{(n)}(t,t_0)$, solutions of eq.~(\ref{eq:gevp}), which by virtue of eq.~(\ref{eq:spectral}) can be written as 
\begin{equation} 
\begin{split} 
m_{D_q^{(n)}}^{\rm eff}(t)={\rm arccosh}\left[ {\lambda^{(n)}(t+1,t_0) + \lambda^{(n)}(t-1,t_0)\over 2 \lambda^{(n)}(t,t_0)}
\right].
\end{split}
\end{equation}
For the case of the lowest lying state ($n=1$) and the first radial excitation ($n=2$) the signals are illustrated in Fig.~\ref{fig:1}. In the plateau region, each $m_{D_q^{(\prime)}}^{\rm eff}(t)$ is then fitted to a constant $m_{D_q^{(\prime)}}$. 
We checked that the results for the first radial excitation remain stable when we change the size of the matrix of correlators. We also checked that from this study we cannot extract a signal for the second radial excitation: only a few points at $t\lesssim 5$ can be seen before the error bars become overwhelmingly large.   
We also checked that the choice of $t_0$ in the GEVP~(\ref{eq:gevp}) does not make any impact on the results presented here. Finally, we also note that the radial excitations extracted on the plateaux of subtracted correlation functions~(\ref{eq:subtr}) are completely consistent with those obtained from the solution to the GEVP~(\ref{eq:gevp}). 

Concerning the decay constants they are extracted from the matrix element obtained by using eq.~(\ref{eq:me}), with $O_L=P_5$, and  the definition 
\bea
(\mu_c+\mu_q) \langle 0\vert P_5\vert D_q^{(\prime )}\rangle = m_{D_q^{(\prime )}}^2 f_{D_q^{(\prime )}}\,.
\eea
The fitting intervals for extracting the masses and decay constants for the lowest lying states are 
\begin{align}
&t/a\in [8,22]_{\beta=3.8},&& t/a\in [8,22]_{\beta=3.9},\cr
&t/a\in [12,26]_{\beta=4.05},&& t/a\in [14,30]_{\beta=4.2},\nn
\end{align}
while for the radially excited states the following fit intervals have been chosen,
\begin{align}
&t/a\in [6,10]_{\beta=3.8},&& t/a\in [9,12]_{\beta=3.9},\cr
&t/a\in [9,12]_{\beta=4.05},&& t/a\in [11,14]_{\beta=4.2}.\nn
\end{align}
For some values of the sea quark mass we have a few more points to fit but globally the time intervals noted above are used to obtain the results that we present in tabs.~\ref{tab:02} and ~\ref{tab:03}, in lattice units and for each of the lattice setups employed in this work. 

\subsection{Re-evaluation of $f_{D_s}$ and $f_{D_s}/f_D$}

The results of ref.~\cite{etmc-D} included the simulations at three different lattice spacings and the value $f_{D_s}=244(8)$~MeV has been reported. That value has been improved in ref.~\cite{etmc-B} where the simulations at a smaller lattice spacing have been included in the analysis, leading to   $f_{D_s}=248(6)$~MeV. Furthermore, while improving the MtmQCD estimate of $f_{D_s}/f_D$, the authors of ref.~\cite{etmc-B} also added the systematic uncertainty related to the chiral extrapolation, which was omitted in ref.~\cite{etmc-D}. Their final result, $ f_{D_s}/f_D = 1.17(5)$, allowed to deduce $f_D=212(8)$~MeV.

Results in refs.~\cite{etmc-D,etmc-B} have been obtained from the correlation functions with local source operators only. In the present work we implement several levels of the smearing procedure discussed above, with parameters~(\ref{eq:params}), and then combine the resulting correlators in a matrix. Solution to the GEVP, together with a slightly modified procedure to extract $f_{D_s}$ and $f_{D_s}/f_D$, result in more accurate results which is why in this subsection we update the values presented in refs.~\cite{etmc-D,etmc-B}. 
 \begin{figure*}[t!]
 \epsfig{file=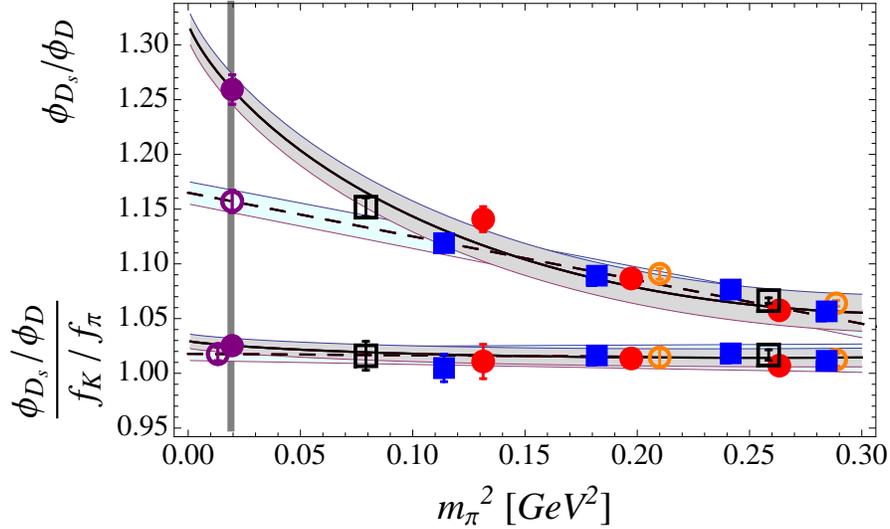, height=7.cm}\\
\caption{\label{fig:2}{\sl 
Chiral extrapolation of the ratio of decay constants: Upper curves correspond to the fit of our data for $\phi_{D_s}/\phi_D$ according to eq.~(\ref{eq:fform-2}) with $X=0$ and $X=1$; Lower curves correspond to the chiral extrapolation 
of the double ratio $(\phi_{D_s}/\phi_D)/(f_K/f_\pi)$ with the formula~(\ref{eq:fform-R}) with $X=0$ and $X=1$. The vertical line indicates the physical pion mass. Dashed (solid) curves depict the extrapolation without (with) inclusion of the chiral logarithms. 
The symbols corresponding to the lattice data are: {\Large\color{BurntOrange}\bf{\textopenbullet}} for $\beta=3.80$, ${\color{blue} \large \blacksquare}$ for $\beta=3.90$,  {\color{red} \LARGE \textbullet} for $\beta=4.05$,  and  ${\color{black} \bm{ \square}}$ for $\beta=4.20$. Note that the result of the linear extrapolation of the double ratio has been slightly off-set to make it distinguishable from the one in which the chiral logarithms have been included.} }
\end{figure*}

To get the physically relevant $f_{D_s}$ we need to extrapolate the values we obtained from all of our lattice ensembles (cf. Tab.~\ref{tab:03}). We choose to combine $f_{D_s}$ and $m_{D_s}$ in the dimensionless ratio that we then fit 
to the form, 
\begin{align}\label{eq:fform-1}
\left({f_{D_s}\over m_{D_s}} \right)^{\rm latt.}\!\!\!=A_{D_s}  \left[ 
1 + B_{D_s} m_q + C_{D_s} \left( {a\over a_{\beta=3.9}}\right)^2
\right],
\end{align}
where $A_{D_s}$, $B_{D_s}$ and $C_{D_s}$ are obtained from the fit and their values are given in Tab.~\ref{tab:04}. The above form takes into account the fact that the lattice discretization effects to the hadronic quantities computed in MtmQCD are $\propto a^2$. Note that we divided by the lattice spacing obtained at $\beta=3.9$ so that the parameter $C_{D_s}$ actually indicates a size of discretization effects at $\beta=3.9$.  
After taking $m_q\equiv m_q^\msbar (2\,\gev)$, also listed in Tab.~\ref{tab:02}, the fit of our data to eq.~(\ref{eq:fform-1}) in the continuum limit and at the physical $m_{u,d}^\msbar (2\,\gev)=3.6(2)$~MeV~\cite{Blossier:2010cr},  gives 
\bea
&& \left({f_{D_s}\over m_{D_s}}\right)^{\rm ph.} \!\! = 0.1281(11)\,.
\eea
With the help of $m_{D_s}^{\rm ph.} = 1968.5(3)$~MeV~\cite{PDG}, we finally have
\bea\label{eq:fDS}
f_{D_s}= 252(3)\ \mev\,.
\eea
We checked that this result remains stable if we omit from the continuum extrapolation the results obtained at $\beta=3.8$. This result is also consistent with those obtained from simulations with $\nf=2+1$ flavors of staggered quarks in the continuum limit~\cite{stagg}, with those computed with $\nf=2+1$ flavors of Wilson-Clover quarks at the single lattice spacing~\cite{Namekawa:2011wt}, as well as with the recent experimental results presented in ref.~\cite{fDsEXP}.

As for the SU(3) light flavor breaking, the ratio of $f_{D_s}/f_D$ is combined with the meson masses in the way consistent with the heavy quark expansion, 
\bea
r_q={\phi_{D_s}\over \phi_{D_q}}\equiv {\sqrt{m_{D_s}\over m_{D_q}} \ {f_{D_s}\over f_{D_q}} }\,, 
\eea
where the index ``$q$" labels the valence light quark, which in our study is mass degenerate with the sea quark. As in ref.~\cite{etmc-D,etmc-B} we fit our results to a form 
\begin{equation} \label{eq:fform-2}
\begin{split} 
r_q^{\rm latt.}= A_r &\left[  1 + X  {3\over 4} {1+3 g^2\over (4\pi f)^2} m_\pi^2 \log(m_\pi^2)  \right. \cr
&\left. \quad\, + B_{r} m_\pi^2 + C_{r} \left( {a\over a_{\beta=3.9}}\right)^2
\right],
\end{split}
\end{equation}
where for $X=0$ we have the expression similar to the one used in eq.~(\ref{eq:fform-1}), and for $X=1$ the extrapolation formula includes the chiral logarithmic correction that has been computed in the framework of heavy meson chiral perturbation theory~\cite{HMChPT}. To use the latter formula one needs to fix the value of the soft pion coupling to the doublet of the lowest lying heavy-light mesons, $g$. That coupling has been recently computed in ref.~\cite{our-g} on the same sets of gauge field configurations that are used here, and the result is $g=0.53(3)(3)$. The results of the fit of our data to eq.~(\ref{eq:fform-2}) are collected in Tab.~\ref{tab:04}. Here we note that
\bea\label{eq:av0}
{\textrm{ for }} X=0,&&  {f_{D_s}\over f_D}= 1.128(10),\nn\\
                                  && \nn\\
{\textrm{ for }} X=1,&& {f_{D_s}\over f_D}= 1.227(13).
\eea
After averaging the last two results, we finally have 
\bea\label{eq:av1}
{f_{D_s}\over f_D}= 1.177(13)(50)\,,
\eea
from which we can deduce $f_D= 214(4)(9)$~MeV, where the second error reflects the systematics arising from the chiral extrapolation.  
In order to circumvent the large logarithmic correction in eq.~(\ref{eq:fform-2}), one can study a double ratio~\cite{sdsj},
\bea
R_q=\sqrt{m_{D_s}\over m_{D_q}} \ {f_{D_s}/f_{D_q} \over f_{K_q}/f_{\pi_{qq}}}\,, 
\eea
for which the logarithmic term is about $10$ times smaller than in eq.~(\ref{eq:fform-2}),
\begin{equation} \label{eq:fform-R}
\begin{split} 
R_q^{\rm latt.}= A_R &\left[  1 + X  \  {9 g^2-2\over 4 \ (4\pi f)^2} m_\pi^2 \log(m_\pi^2)  \right. \cr
&\left. \quad\, +B_{R} m_\pi^2 + C_{R} \left( {a\over a_{\beta=3.9}}\right)^2
\right],
\end{split}
\end{equation}
and therefore the difference between the values obtained by setting $X=1$ and $X=0$ is much smaller, which can also be appreciated from the plot shown in Fig.~\ref{fig:2}. We get 
\bea\label{eq:DRR}
{f_{D_s}\over f_D}= 0.995(6)(4)\times {f_K\over f_\pi}\,,
\eea 
where the central value is obtained by averaging the results of extrapolations with $X=0$ and $X=1$, and the second error reflects the error due to chiral extrapolation. The results of the fit of our data to eq.~(\ref{eq:fform-R}) are listed in Tab.~\ref{tab:04}. 
Following the same strategy described in ref.~\cite{etmc-D}, from the results for $f_{K_q}/f_{\pi_{qq}}$ listed in Tab.~\ref{tab:03} we obtain $f_K/f_\pi= 1.23(1)$, which then gives~\footnote{In addition to the results considered in ref.~\cite{etmc-D}, in this analysis we also included the values of $f_{K_q}/f_{\pi_{qq}}$ obtained at $\beta=4.20$.}
\bea\label{su3}
{f_{D_s}\over f_D} = 1.23(1)(1)\,,
\eea
that combined with $f_{D_s}$ in eq.~(\ref{eq:fDS}) gives
\bea\label{eq:fD}
f_D= 205(5)(2)~\mev .
\eea 
We note also that the above result remains remarkably stable if the data on our coarser lattices (corresponding to $\beta=3.8$) are left out from the chiral and continuum extrapolation.

\subsection{Ratios $f_{D_s^\prime}/f_{D_s}$ and $f_{D^\prime}/f_D$}
\begin{figure*}[t!]
 \epsfig{file=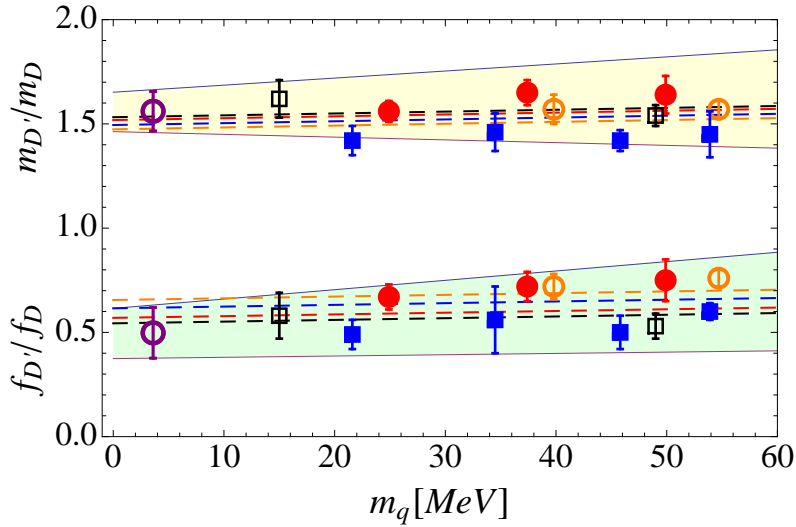, height=7.cm}\\
\caption{\label{fig:3}{\sl 
Chiral extrapolation of the ratios ${m_{D^\prime}/m_{D}}$ and  ${f_{D^\prime}/f_{D}}$ by using eq.~(\ref{eq:fform-F}). Dashed lines correspond to central values of various lattice spacings, and the bands indicate the error bars of extrapolation. The symbols of the lattice data points are the same as in Fig.~\ref{fig:2}.} }
\end{figure*}

We now discuss the masses and the decay constants of the radially excited $D$-mesons. We focus to the dimensionless $m_{D^\prime_q}/m_{D_q}$ and $f_{D^\prime_q}/f_{D_q}$, that are easily built from our results presented in Tab.~\ref{tab:02} ({\sl non-strange}) and Tab.~\ref{tab:03} ({\sl strange}). 
In the following we denote by ${\cal F}$ one of the four quantities discussed in this section, namely $m_{D^\prime}/m_{D}$,  $m_{D^\prime_s}/m_{D_s}$, $f_{D^\prime}/f_{D}$,  and $f_{D^\prime_s}/f_{D_s}$, and fit each to the form similar to eq.~(\ref{eq:fform-1}), 
\begin{align}\label{eq:fform-F}
{\cal F}^{\rm latt.} = A_{\cal F} \left[ 
1 + B_{{\cal F}} m_q + C_{{\cal F}} \left( {a\over a_{\beta=3.9}}\right)^2
\right],
\end{align}
We get the following physically relevant results, 
\bea\label{eq:excited}
&& {m_{D_s^\prime}\over m_{D_s}} = 1.53(7),\quad   {f_{D_s^\prime}\over f_{D_s}} = 0.53(9),\nn\\
&&\nn\\
&&  {m_{D^\prime}\over m_{D}} = 1.56(9),\quad   {f_{D^\prime}\over f_{D}} = 0.50(12).
\eea
An illustration of that fit in the case of ${m_{D^\prime}/m_{D}}$ and ${f_{D^\prime}/f_{D}}$ is provided in Fig.~\ref{fig:3}, while the values of $A_{{\cal F}}$, $B_{{\cal F}}$, and $C_{{\cal F}}$ for all four quantities can be found in Tab.~\ref{tab:04}. 
We observe that the above ratios do not exhibit a regular behavior in $a^2$, and are practically independent of the light quark mass.  
For that reason it is tempting to fit our data to a constant, i.e. to impose $B_{{\cal F}}= C_{{\cal F}} =0$ in eq.~(\ref{eq:fform-F}). We obtain
\bea\label{eq:excited2}
&& {m_{D_s^\prime}\over m_{D_s}} = 1.52(2),\quad   {f_{D_s^\prime}\over f_{D_s}} = 0.65(3),\nn\\
&&\nn\\
&&  {m_{D^\prime}\over m_{D}} = 1.56(9),\quad   {f_{D^\prime}\over f_{D}} = 0.65(3).
\eea
In other words the mass ratios remain stable while the ratios of decay constants change quite considerably. We decide to take the difference between the central values in eq.~(\ref{eq:excited}) and in eq.~(\ref{eq:excited2}) as an estimate of systematic uncertainty and after symmetrizing the error bars we finally obtain:
\bea\label{eq:excitedF}
&& {m_{D_s^\prime}\over m_{D_s}} = 1.53(7),\quad   {f_{D_s^\prime}\over f_{D_s}} = 0.59(11),\nn\\
&&\nn\\
&&  {m_{D^\prime}\over m_{D}} = 1.55(9),\quad   {f_{D^\prime}\over f_{D}} =0.57(16).
\eea

With respect to the mass of the state suggested to be interpreted as $D^\prime$ and measured at BaBar, $m_{D^\prime}/m_D= 1.36$, our result is si\-gn\-ificantly larger. If the state measured at BaBar is indeed the radial excitation $D^\prime$, the fact that our value is larger would be difficult to explain. 

One could suspect that tuning the twisting angle to its maximal value on the lattice could be made only up to discretization effects which then induce a pollution to the extraction of the desired hadron state by the state with opposite parity. However, since that pollution is an ${\cal O}( a^2)$ effect~\cite{fr} and since we perform the extrapolation to the continuum limit, that argument could not be used to explain the potential discrepancy between our findings and the value measured at BaBar. 

To further study this issue, we used the data at fixed lattice spacing (corresponding to $\beta=4.05$) and compared them with the results obtained by using the Wilson-Clover quark action with $\nf=2$ light flavors at nearly equal lattice spacing.~\footnote{The gauge field configurations with Wilson-Clover action have been produced by the QCDSF Collaboration~\cite{qcdsf} that we have used in our previous work~\cite{with-haas}. See also Tab.~\ref{tab:01bis}.} The results are 
\bea
{\rm MtmQCD}:&& {m_{D_s^\prime}\over m_{D_s}} = 1.55(6),\quad   {f_{D_s^\prime}\over f_{D_s}} = 0.69(5),\nn\\
{\rm Clover}:&& {m_{D_s^\prime}\over m_{D_s}} = 1.48(7),\quad   {f_{D_s^\prime}\over f_{D_s}} = 0.77(9).\nn
\eea
In other words, at fixed lattice spacing and fixed light sea quark mass [$ m_q^\msbar (2\,\gev)\simeq 25$~MeV], the results for ${m_{D_s^\prime}/m_{D_s}}$ and ${f_{D_s^\prime}/f_{D_s}}$  obtained by using MtmQCD action are consistent with those obtained with the Wilson-Clover action. We therefore conclude that, within the above error bars, the mass of the radial excitation obtained on the lattice with $\nf=2$ dynamical quarks is larger than the state measured by BaBar. 

Another potential difficulty when computing the properties of higher excited
states on the lattice with light dynamical quarks is that the decay channels
with emission of a pion might open up and modify the value of extracted mass and
decay constant. In the problem at hand, such channels are $D^\prime \to
D^\ast\pi$ and/or $D^\prime \to D_0^\ast \pi$ in the case of non-strange radial
excitation, and $D_s^\prime \to D^\ast K$ or $D_s^\prime \to D_{s0}^\ast K$ for
the strange radial excitation. That difficulty does not exist in quenched QCD
($\nf=0$) which is why we produced a set of quenched QCD  configurations at a similar lattice spacing (cf. Tab.~\ref{tab:01bis}) and computed the mass and decay constant of $D_s$ meson and its radial excitation by following the same steps as described above. We have
\bea
{\rm \nf=0}:&& {m_{D_s^\prime}\over m_{D_s}} = 1.41(9),\quad   {f_{D_s^\prime}\over f_{D_s}} = 0.67(12),\nn\\
{\rm \nf=2}:&& {m_{D_s^\prime}\over m_{D_s}} = 1.48(7),\quad   {f_{D_s^\prime}\over f_{D_s}} = 0.77(9).\nn
\eea  
From our data we cannot see the effects of the sea quark mass. 
We therefore conclude that within the statistical errors of this study the radially excited $D^\prime$ state is heavier than the one measured by BaBar as well as the one predicted by the constituent quark model of ref.~\cite{gi}.  It should be emphasized that this conclusion is based on the comparison made for the case of the strange valence light quark. We did not explore the lighter quarks to be able to make stronger statement for the non-strange $m_{D^\prime}/m_D$. We believe more research is needed in that direction, to explore the simulations with very light sea and valence quarks and check whether or not the dependence of $m_{D^\prime}$ on the light quark mass changes considerably when close to the chiral limit, similar to the findings of ref.~\cite{roper} for the Roper resonance. An indication that this indeed could be the case is provided by the results reported in ref.~\cite{mohler}, where the simulations at one lattice spacing have been used to compute the spectrum of $D_{(s)}$-mesons.~\footnote{Very recently two preliminary studies appeared in ref.~\cite{prelim-D}. They do not report the numerical values for $m_{D_{(s)}^\prime }$ but from the plots provided we could see that their $m_{D^\prime}$ is  larger than the mass of the state observed at BaBar, $2538(8)$~MeV.}

Concerning our results for the decay constants of the radial excitations, we see from eq.~(\ref{eq:excited}) 
that they are considerably smaller than the those of the lowest states. 
This situation is qualitatively different from what happens in the heavy quark limit ($m_c\to \infty$), in which~\cite{burch} 
\bea
\lim_{m_c\to \infty}\left({\sqrt{m_{D_q^\prime}}\ f_{D_q^\prime}
\over \sqrt{m_{D_q}}\  f_{D_q}
}\right) >1\,, 
\eea
while in our case, with the propagating charm quark, the above ratio is smaller than one.

\section{More phenomenology\label{sec:Pheno}}

With the results of our lattice study given in eqs.~(\ref{eq:fD},\ref{eq:excited}), we now return to eq.~(\ref{eq:three}) in which the only remaining ingredient we need is the ratio of Wilson coefficients $a_2/a_1$. Their values have been computed through 
matching of the full Standard Model with the low energy effective theory in which the properly resummed next-to-leading logarithmic QCD corrections have been included. The final result in the $\msbar$(NDR) scheme at $\mu=m_b$ is 
$a_2/a_1= 0.172$~\cite{buras}. 
In the factorization approximation, however, $a_2/a_1$ is considered to be an effective parameter that can be extracted from similar non-leptonic decay modes~\cite{neubert,fact2,a2a1}. We can use the corresponding $B\to D\pi$ decays, and from
\begin{equation}
\begin{split} 
{{\cal B}( B^-\to D^{ 0}\pi^-)\over {\cal B}(\overline B^0\to  D^{+}\pi^-)} &={\displaystyle{\tau_{B^-}\over \tau_{\bar B^0}}}\left[ 1 + \left({a_2\over a_1}\right)^{\rm eff.} {m_B^2-m_\pi^2\over m_B^2-m_{D}^2} \right. \cr
&\hspace*{11mm}\left. \times  {f_0^{B\to \pi}(m_{D}^2)\over f_+^{B\to D}(0)}\ {f_D\over f_\pi}\right]^2\,,
\end{split}
\end{equation}
we get 
\bea
\left({a_2\over a_1}\right)^{\rm eff.}  = 0.368(65)\,.
\eea
To get the last number we used the following input values: (i) $ {\cal B}(\overline B^0\to  D^{+}\pi^-) = 0.286(13)\%$, ${\cal B}( B^-\to D^{ 0}\pi^-)= 0.481(15)\%$~\cite{PDG}, (ii) $\tau_{\bar B^0}/\tau_{B^-}=1.079(7)$~\cite{PDG}, (iii) $f_+^{B\to D}(0)= 0.64(2)$~\cite{ffatzero}, (iv) $f_0^{B\to \pi}(m_D^2)= 0.29(4)$~\cite{blazenka}, (v)  $f_D/f_\pi=1.56(3)(2)$, obtained in this work, cf. eq.~(\ref{eq:excitedF}).

Therefore, with our $f_{D^\prime}/f_D=0.57(16)$ inserted in eq.~(\ref{eq:three}), we have
\bea\label{eq:XYZ}
{{\cal B}( B^-\to D^{\prime 0}\pi^-)\over {\cal B}(\overline B^0\to  D^{\prime +}\pi^-)} &={\displaystyle{\tau_{B^-}\over \tau_{\bar B^0}}} \left[ 1 + \displaystyle{0.14(4)\over f_+^R(0)} \right]^2\,,
\eea
where we used our result  $m_{D^\prime}/m_D= 1.55(9)$. If, instead, we used $m_{D^\prime}/m_D= 1.36$ as
suggested by BaBar, the numerator on the r.h.s. side would change to $0.13(4)$, as quoted in eq.~(\ref{eq:C3R}). 

In addition, by using our value for $m_{D^\prime}/m_D$ the result of eq.~(\ref{eq:c1}) would become
\bea
{{\cal B}(\overline B^0\to D^{\prime +}\pi^-)\over {\cal B}(\overline B^0\to D^{+}\pi^-)} =
(1.24\pm 0.21)\times |f_+^R(0)|^2\,,
\eea
which is to be compared with~(\ref{eq:CC}). We see that both ratios~(\ref{eq:c1},\ref{eq:three}) do not substantially depend on the mass of the radial excitation, and the measurement of the decay of $B$-meson to the radially excited $D^\prime$ should be feasible if the form factor $f_+^R(0)$ is significantly enhanced by the power corrections.

To further emphasize the experimental feasibility of measuring ${\cal B}(B\to D^\prime \pi)$ decays for large $f_+^R(0)$, we can use $f_+^R(0)=0.4$ as obtained in ref.~\cite{galkin} and compare this decay to ${\cal B}(B\to D_2^\ast \pi)$ 
that was measured at Belle and BaBar in both cases, Class-I~\cite{B0D2pi} and  Class-III~\cite{BmD2pi}. We get
\begin{align}\label{eq:51}
&{{\cal B}( \overline B^0\to  D^{\prime +}\pi^-)\over {\cal B}(\overline B^0\to  D_2^{\ast +}\pi^-)} =1.6(3) && [1.2(3)], \nn\\
&{{\cal B}( B^-\to D^{\prime 0}\pi^-)\over {\cal B}(B^- \to  D_2^{\ast 0}\pi^-)} =1.4(3)&& [1.1(3)],
\end{align}  
where the first number on the r.h.s. corresponds to $m_{D^\prime}/m_D=1.36$ and the number within the brackets to $m_{D^\prime}/m_D=1.55(9)$.  
In other words the measurement of ${\cal B}(B\to D^\prime \pi)$ is as feasible as that of ${\cal B}(B\to D_2^\ast \pi)$, provided the form factor $f_+^R(0)$ is large as recently suggested in the literature. 
Note also that for the above estimates we used  ${\cal B}(\overline B^0\to  D_2^{\ast +}\pi^-)=4.9(7)\times 10^{-4}$, and ${\cal B}(B^- \to  D_2^{\ast 0}\pi^-)=8(1)\times 10^{-4}$~\cite{puzzle1}.
Experimentally, these measurements are made in the Dalitz plot analysis of $B\to D^\ast \pi \pi$ decay. It is therefore important to mention that the statement made in eq.~(\ref{eq:51}) remains valid even after we take into account the results of refs.~\cite{swanson,goity}, namely ${\cal B}(D^\prime \to D^\ast \pi)\gtrsim {\cal B}(D_2^\ast \to D^\ast \pi)$. In other words, for a large value of the form factor $f_+^R(0)$ the decay mode $B\to  D^{\prime }\pi$ should be as discernible in the Dalitz plot analysis of $B\to D^\ast \pi \pi$ as $B\to  D_2^{\ast}\pi$.

\section{Summary\label{sec:Summary}}
In this paper we develop the possibilities to check experimentally the size of the form factor $f_+^R(q^2)$ that parameterizes the $B\to D^\prime$ weak transition matrix element~\cite{bernlochner}. If its value is large, as recently claimed, the decay $\overline B^0\to D^{\prime +}\pi^-$ should be accessible experimentally and we predict its branching fraction in terms of $|f_+^R(0)|^2$, by using the measured ${\cal B}(\overline B^0\to D^{+}\pi^-)$ and by employing the factorization approximation to describe the decay amplitudes. The uncertainties to the used approximation are expected to be small since the ratios of the measured Class-I non-leptonic decays are known to be very well described by factorization. 

If the experimenters succeed in measuring ${\cal B}(\overline B^0\to D^{\prime +}\pi^-)$, it would be interesting to  check whether or not the corresponding $|f_+^R(0)|^2$ is consistent with the ratio between the Class-III and Class-I $B\to D^\prime \pi$ decay modes (\ref{eq:XYZ}), which we could also predict using the  factorization approximation. Note that the factorization in Class-III decays requires more assumptions as well as the computation of the decay constant $f_{D^\prime}$. 

We computed $m_{D^\prime}/m_D$ and $f_{D^\prime}/f_D$ by using the gauge field configurations with $\nf=2$ mass-degenerate light quark flavors, generated at four lattice spacings and for several light sea quark masses. 
We find 
\bea
{m_{D^\prime}\over m_D}=1.55(9),\quad {f_{D^\prime}\over f_D}=0.57(16).
\eea 
If the state observed by BaBar Collaboration is indeed $D^\prime$, then our result is larger than theirs, $m_{D^\prime}/m_D=1.36$. More research on both sides is needed to clarify the (potential) discrepancy. On the lattice QCD side it would be interesting to check whether or not $m_{D^\prime}/m_D$ becomes sensitive to the variation of the light quark mass in the region with very light quarks (closer to the chiral limit), the region not explored in the present study. Such a situation, that a hadron mass strongly depends on the sea quark mass when the latter is close to the chiral limit, was observed in the case of the Roper resonance on the lattice~\cite{roper}.  Concerning the interpretation of the state observed by BaBar at $2539(8)$~MeV, it is important to understand why its width is much larger than predicted. As a starting point one could verify if the predictions of ref.~\cite{swanson} remain stable if one uses different set of wave functions (for example those of the model of ref.~\cite{gi}) or different models. 

We also improved the computation of the decay constants $f_{D_{(s)}}$ by relying on the chiral and continuum extrapolation of the ratios  $f_{D_s}/m_{D_s}$ and $(\phi_{D_s}/\phi_D)/(f_K/f_\pi)$. More specifically we obtain:
\bea
f_{D_s}= 252(3)~\mev ,\quad {f_{D_s}\over f_D}= 1.23(1)(1),
\eea
where the second error in the latter result reflects the uncertainty due to inclusion/omission of the chiral logarithms in the light mass extrapolation to the physical limit. These two results give $f_D=205(5)(2)$~MeV.~\footnote{Recent estimates of the vector meson decay constants  $f_{D_s^\ast}$, $f_{D_s^\ast }/f_{D^\ast}$, can be found in ref.~\cite{vector}.} 

Finally, our prediction for the non-leptonic decays depends only mildly on the choice of the mass of the radial excitation. For $m_{D^\prime}/m_D=1.36$ we have
\begin{equation} 
\begin{split} 
{\cal B}(\overline B^0\to &D^{\prime +}\pi^-) = [4.7(4)\times 10^{-3}] \  |f_+^R(0)|^2, \cr
{\cal B}( B^-\to &D^{\prime 0}\pi^-) =\cr
&[5.1(5)\times  10^{-3}] \left[ 1 + \displaystyle{0.13(4)\over f_+^R(0)} \right]^2  |f_+^R(0)|^2 ,
\end{split}
\end{equation}
and for $m_{D^\prime}/m_D=1.55(9)$ we get
\begin{equation} 
\begin{split} 
{\cal B}(\overline B^0\to &D^{\prime +}\pi^-) = [3.6(6)\times 10^{-3}] \  |f_+^R(0)|^2, \cr
{\cal B}( B^-\to &D^{\prime 0}\pi^-) =\cr
&[3.8(7)\times  10^{-3}] \left[ 1 + \displaystyle{0.16(5)\over f_+^R(0)} \right]^2  |f_+^R(0)|^2 .
\end{split}
\end{equation}
\section*{Acknowledgements}
We thank Luis Oliver for discussions about the phenomenological aspects of $B$ decays to excited $D$-mesons. F.S. thanks Paula Rubio for a helpful discussion on the determination of excited states.  We thank the ETMC for making their gauge field configurations publicly available.
Computations of the relevant correlation functions are made on GENCI-IDRIS, under the Grant 2012-056806. 

\begin{table*}[h!!]
\begin{ruledtabular}
\begin{tabular}{|c|cccccc|}   
{\phantom{\huge{l}}}\raisebox{-.2cm}{\phantom{\Huge{j}}}
$ \beta$& 3.8 &  3.9  &  3.9 & 4.05 & 4.2  & 4.2    \\ 
{\phantom{\huge{l}}}\raisebox{-.2cm}{\phantom{\Huge{j}}}
$ L^3 \times T $&  $24^3 \times 48$ & $24^3 \times 48$  & $32^3 \times 64$ & $32^3 \times 64$& $32^3 \times 64$  & $48^3 \times 96$  \\ 
{\phantom{\huge{l}}}\raisebox{-.2cm}{\phantom{\Huge{j}}}
$ \#\ {\rm meas.}$& 240 &  240 & 150 & 150 & 150 & 100  \\ \hline 
{\phantom{\huge{l}}}\raisebox{-.2cm}{\phantom{\Huge{j}}}
$\mu_{\rm sea 1}$& 0.0080 & 0.0040 & 0.0030 & 0.0030 & 0.0065 &  0.0020   \\ 
{\phantom{\huge{l}}}\raisebox{-.2cm}{\phantom{\Huge{j}}}
$\mu_{\rm sea 2}$& 0.0110 & 0.0064 & 0.0040 & 0.0060 &   &     \\ 
{\phantom{\huge{l}}}\raisebox{-.2cm}{\phantom{\Huge{j}}}
$\mu_{\rm sea 3}$&  &   &  & 0.0080 &   &     \\  \hline
{\phantom{\huge{l}}}\raisebox{-.2cm}{\phantom{\Huge{j}}}
$a \ {\rm [fm]}$&   0.098(3) & 0.085(3) & 0.085(3) & 0.067(2) & 0.054(1) & 0.054(1)      \\ 
{\phantom{\huge{l}}}\raisebox{-.2cm}{\phantom{\Huge{j}}}
$\mu_{s}$& 0.0194(7)  &0.0177(6)  &0.0177(6)   & 0.0154(5) & 0.0129(5) & 0.0129(5)  \\ 
{\phantom{\huge{l}}}\raisebox{-.2cm}{\phantom{\Huge{j}}}
$\mu_{c}$& 0.2331(82)  &0.2150(75)  &0.2150(75)   & 0.1849(65) & 0.1566(55) & 0.1566(55)  \\ 
\end{tabular}
{\caption{\footnotesize  \label{tab:01} Lattice ensembles used in this work with the indicated number of gauge field configurations. Lattice spacing is fixed by using the Sommer parameter $r_0/a$~\cite{R0}, with $r_0= 0.440(12)$~fm fixed by matching $f_\pi$ obtained on the lattice with its physical value  (cf. ref.~\cite{Blossier:2010cr}). Quark mass parameters $\mu$ are given in lattice units.}}
\end{ruledtabular}
\end{table*}

\begin{table*}[h!!]
\begin{ruledtabular}
\begin{tabular}{|ccccccc|}   
{\phantom{\huge{l}}}\raisebox{-.2cm}{\phantom{\Huge{j}}}
$\nf$ & $ \beta$ ($c_{SW}$) & $ L^3 \times T $ & $ \#\ {\rm meas.}$ &  $\kappa_{\rm sea}$ & $\kappa_{\rm s}$ & $\kappa_{\rm c}$   \\  \hline
{\phantom{\huge{l}}}\raisebox{-.2cm}{\phantom{\Huge{j}}}
0 & 6.2 (1.614) & $24^3 \times 48$ & 200 &  -- & 0.1348 & 0.125 \\ 
{\phantom{\huge{l}}}\raisebox{-.2cm}{\phantom{\Huge{j}}}
2 & 5.4 (1.823) & $24^3 \times 48$ & 160 & 0.13625 & 0.1359 & 0.126 \\ 
\end{tabular}
{\caption{\footnotesize  \label{tab:01bis} Lattice set-up for the results obtained by using the Wilson gauge and the Wilson-Clover quark action. $\kappa_{\rm sea}$, $\kappa_s$ and $\kappa_c$ stand for the value of the hopping parameter of the sea, strange and the charm quark respectively.}}
\end{ruledtabular}
\end{table*}

\begin{table*}[h!!]
\begin{ruledtabular}
\begin{tabular}{|cc|cccc|} 
{\phantom{\huge{l}}} \raisebox{-.2cm} {\phantom{\huge{j}}}
($L$, $\beta$, $\mu_q$)     & $ m_q^\msbar (2\,\gev)$   &  $ m_{D_q}$  & $ m_{D^\prime_q}$   &   $ f_{D_q}$  & $ f_{D^\prime_q}$               \\ \hline\hline
{\phantom{\huge{l}}} \raisebox{-.2cm} {\phantom{\huge{j}}}
(24, 3.80, 0.0080)                &  0.0398(11) & 0.843(1)      & 1.32(6) & 0.136(1)  & 0.098(8) \\  
{\phantom{\huge{l}}} \raisebox{-.2cm} {\phantom{\huge{j}}}          
 (24, 3.80, 0.0110)               & 0.0547(15)  & 0.852(1)      & 1.34(3) & 0.139(1)  & 0.105(6) \\ \hline
{\phantom{\huge{l}}} \raisebox{-.2cm} {\phantom{\huge{j}}}          
(32, 3.90,0.0040)                 & 0.0216(5)    & 0.741(1)       & 1.05(5) & 0.110(1)  & 0.054(8) \\ 
{\phantom{\huge{l}}} \raisebox{-.2cm} {\phantom{\huge{j}}}          
(24, 3.90,0.0064)                 & 0.0345(8)    & 0.7748(1)    & 1.09(7) & 0.112(1)  & 0.063(18) \\  
{\phantom{\huge{l}}} \raisebox{-.2cm} {\phantom{\huge{j}}}          
(24, 3.90,0.0085)                 & 0.0458(11)  & 0.748(2)      & 1.06(4) & 0.113(1)  & 0.056(9) \\
{\phantom{\huge{l}}} \raisebox{-.2cm} {\phantom{\huge{j}}}          
(24, 3.90,0.0100)                 & 0.0539(13)  & 0.755(1)      & 1.10(3) & 0.116(1)  & 0.069(5) \\ \hline
{\phantom{\huge{l}}} \raisebox{-.2cm} {\phantom{\huge{j}}}          
(32, 4.05,0.0030)                 & 0.0162(4)    & 0.608(2)       & 0.95(3) & 0.083(1)  & 0.055(5) \\ 
{\phantom{\huge{l}}} \raisebox{-.2cm} {\phantom{\huge{j}}}          
(32, 4.05,0.0060)                 & 0.0216(5)    & 0.616(1)       & 1.02(4) & 0.087(1)  & 0.063(6) \\ 
{\phantom{\huge{l}}} \raisebox{-.2cm} {\phantom{\huge{j}}}          
(32, 4.05,0.0080)                 & 0.0249(7)    & 0.621(1)       & 1.02(6) & 0.090(1)  & 0.068(9) \\ \hline
{\phantom{\huge{l}}} \raisebox{-.2cm} {\phantom{\huge{j}}}          
(32, 4.20,0.0065)                 & 0.049(2)      & 0.521(1)       & 0.79(3) & 0.071(1)  & 0.038(4) \\ 
{\phantom{\huge{l}}} \raisebox{-.2cm} {\phantom{\huge{j}}}          
(48, 4.20,0.0020)                 & 0.0150(7)    & 0.497(1)       & 0.81(5) & 0.064(1)  & 0.037(7) \\ 
\end{tabular}

{\caption{  \label{tab:02} \sl
Masses and decay constants, $m_{D_q^{(\prime)}}$ and $f_{D_q^{(\prime)}}$,  as computed from the solution to the GEVP discussed in the text. Note that the light valence quark and the sea quarks are degenerate in mass, $m_q$, 
with the renormalized value given in the $\msbar$ scheme. Note that the hadron masses and decay constants are given in lattice units while $ m_q^\msbar (2\,\gev)$ is given in physical units [GeV].  }}
\end{ruledtabular}
\end{table*}

\begin{table*}[h!!]
\begin{ruledtabular}
\begin{tabular}{|c|cccc|c|} 
{\phantom{\huge{l}}} \raisebox{-.2cm} {\phantom{\huge{j}}}
($L$, $\beta$, $\mu_q$)      &  $ m_{D_s}$  & $ m_{D^\prime_s}$   &   $ f_{D_s}$  & $ f_{D^\prime_s}$      & $ f_K/f_\pi$           \\ \hline\hline
{\phantom{\huge{l}}} \raisebox{-.2cm} {\phantom{\huge{j}}}
(24, 3.80, 0.0080)                & 0.8703(8)      & 1.36(7) & 0.1459(9)  & 0.107(13) &  1.075(4) \\  
{\phantom{\huge{l}}} \raisebox{-.2cm} {\phantom{\huge{j}}}          
 (24, 3.80, 0.0110)               & 0.8717(7)      & 1.35(3) & 0.1462(9)  & 0.109(5)  & 1.051(3) \\ \hline
{\phantom{\huge{l}}} \raisebox{-.2cm} {\phantom{\huge{j}}}          
(32, 3.90,0.0040)                    & 0.7708(8)       & 1.09(4) & 0.1206(7)  & 0.062(7)& 1.114(7)  \\ 
{\phantom{\huge{l}}} \raisebox{-.2cm} {\phantom{\huge{j}}}          
(24, 3.90,0.0064)                   & 0.7715(8)    & 1.11(4) & 0.1203(5)  & 0.066(11) & 1.072(2)  \\  
{\phantom{\huge{l}}} \raisebox{-.2cm} {\phantom{\huge{j}}}          
(24, 3.90,0.0085)                 & 0.7963(10)      & 1.09(3) & 0.1199(7)  & 0.064(8)  & 1.057(2) \\
{\phantom{\huge{l}}} \raisebox{-.2cm} {\phantom{\huge{j}}}          
(24, 3.90,0.0100)                   & 0.7713(8)      & 1.12(2) & 0.1214(6)  & 0.074(4)& 1.045(1) \\ \hline
{\phantom{\huge{l}}} \raisebox{-.2cm} {\phantom{\huge{j}}}          
(32, 4.05,0.0030)                    & 0.6344(11)       & 0.99(3) & 0.0923(7)  & 0.063(5)& 1.129(6)  \\ 
{\phantom{\huge{l}}} \raisebox{-.2cm} {\phantom{\huge{j}}}          
(32, 4.05,0.0060)                 & 0.6355(9)       & 1.03(4) & 0.0930(6)  & 0.066(5)  & 1.072(1)   \\ 
{\phantom{\huge{l}}} \raisebox{-.2cm} {\phantom{\huge{j}}}          
(32, 4.05,0.0080)                    & 0.6361(9)       & 1.03(5) & 0.0941(8)  & 0.072(8) & 1.050(2) \\ \hline
{\phantom{\huge{l}}} \raisebox{-.2cm} {\phantom{\huge{j}}}          
(32, 4.20,0.0065)                     & 0.5243(7)       & 0.81(2) & 0.0750(6)  & 0.041(4)&  1.049(2)  \\ 
{\phantom{\huge{l}}} \raisebox{-.2cm} {\phantom{\huge{j}}}          
(48, 4.20,0.0020)                    & 0.5198(4)       & 0.82(2) & 0.0726(3)  & 0.042(3) & 1.134(6)  \\ 
\end{tabular}

{\caption{  \label{tab:03} \sl
Similar as in Tab.~\ref{tab:02} except that the valence quark mass is fixed to the strange quark mass value.  We also list the values of $f_K/f_\pi$ obtained on each lattice (also referred to in the text as $f_{K_q}/f_{\pi_{qq}}$) which are extracted in the same way as in ref.~\cite{etmc-D} and corrected for the small finite volume effects~\cite{finite-V}. }}
\end{ruledtabular}
\end{table*}

\begin{table*}[h!!]
\begin{ruledtabular}
\begin{tabular}{|c|c|ccc|} 
{\phantom{\huge{l}}} \raisebox{-.2cm} {\phantom{\huge{j}}}
Quantity ($Q$)      &  Fit form   &   $A_Q$  & $B_Q$      & $ C_Q$           \\ \hline 
{\phantom{\huge{l}}} \raisebox{-.2cm} {\phantom{\huge{j}}}
 $f_{D_s}/m_{D_s}$     & eq.~(\ref{eq:fform-1}) & 0.1278(11)  & 0.4(2) &  0.21(1) \\  
{\phantom{\huge{l}}} \raisebox{-.2cm} {\phantom{\huge{j}}}          
 $r=\phi_{D_s}/\phi_{D}$     & eq.~(\ref{eq:fform-2}) [$X=0$] & 1.165(10)  & -0.34(3) &  0.006(5) \\  
{\phantom{\huge{l}}} \raisebox{-.2cm} {\phantom{\huge{j}}}          
 $r=\phi_{D_s}/\phi_{D}$     & eq.~(\ref{eq:fform-2}) [$X=1$] & 1.319(14)  & 0.06(3) &  0.002(5) \\  
{\phantom{\huge{l}}} \raisebox{-.2cm} {\phantom{\huge{j}}}          
  $R=r/(f_K/f_\pi)$     & eq.~(\ref{eq:fform-R}) [$X=0$] & 1.018(6)  & -0.013(22) &  -0.002(4) \\  
{\phantom{\huge{l}}} \raisebox{-.2cm} {\phantom{\huge{j}}}          
  $R=r/(f_K/f_\pi)$     & eq.~(\ref{eq:fform-R}) [$X=1$] &1.029(7)  & 0.021(24) &  -0.002(4) \\  
{\phantom{\huge{l}}} \raisebox{-.2cm} {\phantom{\huge{j}}}          
  $m_{D^\prime}/m_{D}$     & eq.~(\ref{eq:fform-F})   & 1.55(9)  & 0.6(1.5) &  -0.04(5) \\  
{\phantom{\huge{l}}} \raisebox{-.2cm} {\phantom{\huge{j}}}          
  $m_{D_s^\prime}/m_{D_s}$     & eq.~(\ref{eq:fform-F})   & 1.53(7)  & 1.0(1.2) &  -0.05(4) \\  
{\phantom{\huge{l}}} \raisebox{-.2cm} {\phantom{\huge{j}}}          
  $f_{D^\prime}/f_{D}$     & eq.~(\ref{eq:fform-F})   & 0.50(12)  & 1.7(5.6) &  0.24(23) \\  
{\phantom{\huge{l}}} \raisebox{-.2cm} {\phantom{\huge{j}}}          
  $f_{D_s^\prime}/f_{D_s}$     & eq.~(\ref{eq:fform-F})   & 0.52(9)  & 1.5(4.6) &  0.19(21) \\
\end{tabular}

{\caption{  \label{tab:04} \sl
Fit results of the quantities computed in this paper on the lattice.  }}
\end{ruledtabular}
\end{table*}

\end{document}